\documentclass[%
reprint,
amsmath,amssymb,
aps,
]{revtex4-2}
\pdfoutput=1

\usepackage{graphicx}
\usepackage{dcolumn}
\usepackage{bm}

\newcommand{\be}{\begin{equation}}
\newcommand{\ee}{\end{equation}}

\begin{document}
	
	\preprint{APS/123-QED}
	
	\title{Delivering Broadband Light Deep Inside Diffusive Media}
	
	\author{Rohin McIntosh$^1$}
	\author{Arthur Goetschy$^2$}
    \author{Nicholas Bender$^3$}
    \author{Alexey~Yamilov$^4$}
	\author{Chia Wei Hsu$^5$}
	\author{Hasan Y\i lmaz$^6$}
	\author{Hui Cao$^1$}
	\affiliation{
		1.  Department of Physics, Yale University, 06511 New Haven, CT, USA\\
  	2.  ESPCI ParisTech, PSL Research University, CNRS, Institut Langevin, F-75005 Paris, France\\
		3.  Department of Applied Physics, Cornell University, 14850 Ithaca, NY, USA\\
        4.  Department of Physics, Missouri University of Science \& Technology, 65409 Rolla, MO, USA\\
		5.  Department of Electrical and Computer Engineering, University of Southern California, 90089 Los Angeles, CA, USA\\
		6.  Institute of Materials Science and Nanotechnology, National Nanotechnology Research Center (UNAM), Bilkent University, 06800 Ankara, Turkey
	}
	
	\date{September 17, 2023}
	
	\begin{abstract}
		Wavefront shaping enables targeted delivery of coherent light into random-scattering media, such as biological tissue, by constructive interference of scattered waves. However, broadband waves have short coherence times, weakening the interference effect. Here, we introduce a broadband deposition matrix that identifies a single input wavefront that maximizes the broadband energy delivered to an extended target deep inside a diffusive system. We experimentally demonstrate that long-range spatial and spectral correlations result in a six-fold energy enhancement for targets containing more than 1500 speckle grains and located at a depth of up to ten transport mean free paths, even when the coherence time is an order of magnitude shorter than the diffusion dwell time of light in the scattering sample. In the broadband (fast decoherence) limit, enhancement of energy delivery to extended targets becomes nearly independent of the target depth and dissipation. Our experiments, numerical simulations, and analytic theory establish the fundamental limit for broadband energy delivery deep into a diffusive system, which has important consequences for practical applications. 
		
	\end{abstract}
	
	\maketitle
	
	\section{Introduction}
	
	Waves propagate diffusively through disordered media, such as biological tissue, clouds, and paint, due to random scattering. This process is deterministic for elastic scattering in static disordered structures, enabling the control of wave propagation by spatial wavefront shaping \cite{Mosk2012, Rotter2017, Cao2022}. For a coherent beam, spatial modulation of its wavefront incident on a random medium can manipulate the interference of scattered waves from different paths, resulting in light focusing \cite{Vellekoop2007, Yaqoob2008, Vellekoop2008_2, Xu2011, Judkewitz2013, Horstmeyer2015, Vellekoop2015}, enhancement of total transmission \cite{Vellekoop2008, Popoff2010, Choi2011, Kim2012, Yu2013, Popoff2014, Gerardin2014, Davy2015, Hsu2017, Yilmaz2019, Bender2020}, or energy deposition deep inside scattering media \cite{Hsieh2010, Xu2011, Choi2013, Cheng2014, Chaigne2014, Liu2015, Sarma2016, Badon2016, Jeong2018, Katz2019, Durand2019, Boniface2019, Horodynski2020, Yang2019, Boniface2020, Lambert2020, Badon2020, Bouchet2021, Lee2023, Bender2022_2}. Such success relies on the coherence time of the input light $\tau_c$ exceeding the diffusion dwell time $\tau_d$ of light inside the multiple-scattering system, enabling scattered waves to remain phase coherent and thereby interfere. However, many applications use partially coherent light, e.g., broadband light from a supercontinuum source or a super luminescent diode. Once $\tau_c < \tau_d$, dephasing among the scattered waves will weaken their interference, diminishing the power of wavefront shaping. Previous studies have shown that the focusing efficiency decreases for broadband light, as the coherence time $\tau_c$ is inversely proportional to the spectral bandwidth $\Delta \omega$ \cite{Beijnum2011, Paudel2013, Andreoli2015, Mounaix2016, Mounaix2017}. Furthermore, the focal targets in various applications, such as detectors for communications and cells or tissues for photothermal therapy or laser microsurgery, contain many wavelength-scale speckles. Simultaneously enhancing the intensities of all speckles within the target by shaping a single wavefront is more difficult than focusing on a single speckle \cite{Hsu2017}. Having to contend with both temporal decoherence and spatial decorrelation, broadband deposition to a large target is significantly more challenging than narrowband focusing on a wavelength-scale target.
	
	Our aim is to deliver broadband light of $\tau_c \ll \tau_d$ to an extended target with dimensions greater than $\lambda$ at a depth well exceeding the transport mean free path $\ell_t$ in a diffusive system. While short-range correlations of scattered waves dictate light focusing to a wavelength-scale target, long-range spatial correlations play an essential role in delivering light to a large target \cite{Hsu2017}. It has been shown theoretically that nonlocal spectral correlations facilitate broadband enhancement of total transmission by wavefront shaping \cite{Hsu2015}. For energy delivery deep into a turbid medium \cite{Bender2022_2}, however, the capabilities of wavefront shaping for broadband energy enhancement and their underlying physics are not known. Open questions include: is it possible to deterministically find a single wavefront that maximizes broadband energy delivery to an extended target deep inside a diffusive system? What is the fundamental limit for the delivered energy and how does it depend on the target depth and bandwidth? 
	
\begin{figure}
		\includegraphics[width=.48\textwidth]{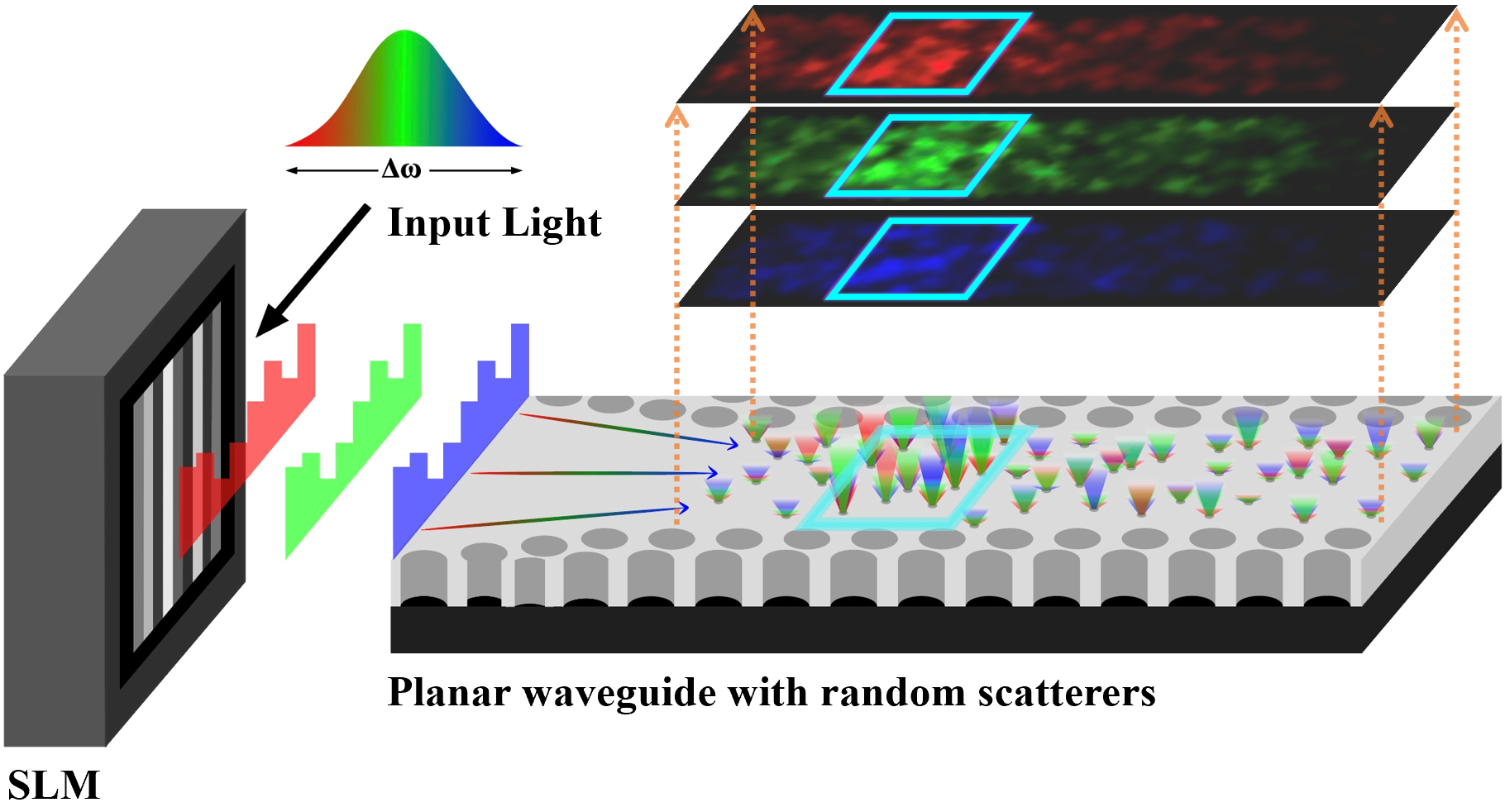}
		\caption{\label{illustration} \textbf{Illustration of experimental platform.} Schematic showing a broadband light is modulated by a spatial light modulator (SLM) and then delivered to an extended target (cyan box) inside a planar waveguide with randomly distributed scatterers. A tapered waveguide segment directs light into the disordered region with length $L$ = 50 \textmu m and width $W$ = 15 \textmu m. The transport mean free path is $\ell_t$ = 3.3 \textmu m. A small fraction of the light is scattered out of the plane, allowing an interferometric measurement of field distribution inside the waveguide. The broadband deposition matrix is constructed experimentally from the frequency-resolved matrix that maps the incident wavefront to the field distribution within the target. Its eigenvector with the largest eigenvalue gives a single spatial wavefront that simultaneously enhances the energy in the target for all frequencies within the input light, as exemplified by three images of intensity distributions at different frequencies.}
	\end{figure}

	To address these questions, we introduce the broadband deposition matrix, $\mathcal{A}$, whose eigenvector gives the input wavefront that maximizes the energy (summed over all input frequencies) delivered to a target of arbitrary size and shape. Our experimental platform is schematically illustrated in Fig.~\ref{illustration}, depicting a single input wavefront that optimizes energy delivery for multiple frequencies. Even when the coherence time of an input light is ten times shorter than the dwell time of diffusive waves in the disordered medium, we observe a six-fold enhancement of total energy (spatially and temporally integrated intensity) in a large target that contains approximately 1700 speckles at a depth of $9 \, \ell_t$. The long-range spectral correlations greatly slow down the drop of energy enhancement with increasing bandwidth (decreasing coherence time $\tau_c$). In addition, unlike the broadband focusing enhancement that decays rapidly with the depth, the broadband energy enhancement over a large target increases slightly with the depth, and becomes nearly depth invariant when $\tau_c \ll \tau_d$. These results can be explained by the distinct depth dependence of short-range correlations for the wavelength-scale target and long-range correlations for the extended target. The sustainability of energy enhancement for broad bandwidths and large areas at any depth illustrates that long-range spatial and spectral correlations are vital for energy delivery deep into diffusive media by wavefront shaping. 
	
	A further question is how dissipation or absorption affects the efficiency of wavefront shaping. Previous studies show that optical dissipation in random media profoundly impacts coherent control of wave transport \cite{Liew2014, Liew2015, Sarma2015, Yamilov2016}. The dissipation strength in a diffusive medium can be described by the dissipation time $\tau_a$. We find that once the coherence time of the input light $\tau_c$ becomes shorter than $\tau_a$, the dissipation effects on the large-area energy enhancement diminish. This is attributed to the fact that the decoherence effect overwhelms the dissipation effect. More specifically, the decoherence of scattered waves is more detrimental to the interference effect than the attenuation of long scattering paths by dissipation. This conclusion is supported by our numerical simulations and analytic theory. 
 
 We note that all the results from this study are generalizable to the linear scattering of any waves, including acoustic and electron waves. In particular, the broadband deposition matrix introduced here provides the fundamental physical limit for single-wavefront energy delivery into diffusive media for any input bandwidth, target area, and deposition depth, giving a general upper bound with which to constrain wavefront optimization.
	
	\section{Broadband Deposition Matrix}
	
	Monochromatic energy deposition can be described by the deposition matrix $\mathcal{Z}(\omega_0)$ which maps incident wavefronts at frequency $\omega_0$ to internal electric fields in a target of arbitrary shape and size \cite{Bender2022_2}. The eigenvector of $\mathcal{Z^\dag}(\omega_0)\mathcal{Z}(\omega_0)$ with the largest eigenvalue gives the incident wavefront of a monochromatic light that deposits maximal energy (spatially integrated intensity at $\omega_0$) to the target region. As the frequency of the input light $\omega$ is detuned from $\omega_0$, the energy deposited by the aforementioned eigenvector decreases rapidly, approaching that of random wavefront illuminations. The full width at half maximum $\delta\omega$ of this decay curve is inversely proportional to the dwell time of light $\tau_d$~\cite{Hsu2015}. Deeper into a diffusive system, $\tau_d$ is longer, and $\delta\omega$ becomes narrower. 
	
	For broadband deposition, we consider an input light of frequency bandwidth $\Delta \omega$. Conventionally, the number of independent spectral channels is $M = {\Delta\omega}/{\delta\omega} + 1$. Each spectral channel requires a distinct wavefront for maximal energy delivery to the same target. To find a single wavefront that maximizes energy delivery over the entire bandwidth of $\Delta\omega>\delta\omega$, we introduce the broadband deposition matrix 
	\begin{equation}
	\mathcal{A} = \int_{\Delta\omega} d\omega \, I(\omega) \, \mathcal{Z^\dag(\omega)}\mathcal{Z(\omega)},
	\label{A}
	\end{equation}
	where $I(\omega)$ is the spectral intensity of input light, normalized to $\int_{\Delta\omega} d\omega \, I(\omega)=1$.  The eigenvector of $\mathcal{A}$ with the largest eigenvalue represents the single wavefront that maximizes the energy deposition over the input bandwidth $\Delta \omega$ in a given target region. The corresponding eigenvalue is the maximal energy (spectrally integrated intensity within $\Delta \omega$) on the target.
	
	\section{Experiment of broadband energy delivery}
	
	Experimentally, we study broadband optical energy delivery into a diffuse waveguide, as schematically depicted in Fig.~\ref{illustration}. The planar waveguides are fabricated on a silicon-on-insulator wafer with reflective photonic-crystal sidewalls to confine light. Randomly arranged air holes with a diameter of $d = 100$~nm are etched into the silicon layer, serving as scatterers. A coherent monochromatic beam from a tunable infrared (IR) laser is shaped by a spatial light modulator (SLM) and then coupled into a silicon waveguide. The waveguide segment with randomly distributed air holes has a length of $L$ = 50 \textmu m and a width of $W$ = 15 \textmu m. It supports $N = 56$ guided modes in the 1551 nm to 1556 nm wavelength range. The air hole density is 5.5\%, and the transport mean free path is $l_t$ = 3.3 \textmu m. Since $L\gg \ell_t$, light undergoes multiple -scattering as it traverses the disordered region of the waveguide. A small amount of light is scattered out of the plane, enabling direct probing of the electric field distribution inside the waveguide. We perform an interferometric measurement of light scattered out of the plane with a reference beam. A CCD camera records their interference pattern. Additional details of the experiment are provided in Appendix A.
	
	\begin{figure}
		\includegraphics[width=.48\textwidth]{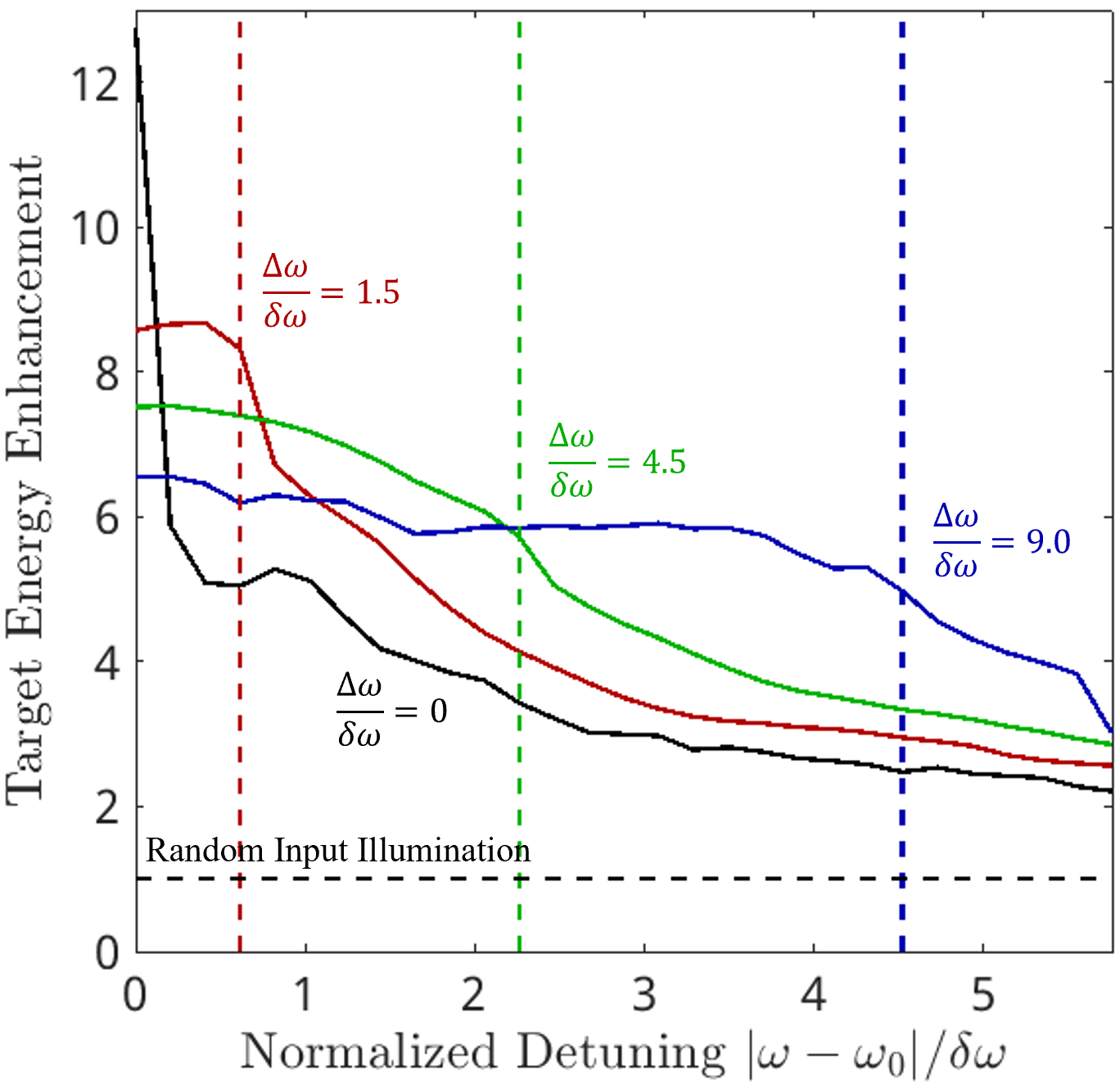}
		\caption{\label{dep_spec} \textbf{Targeted broadband energy delivery.} Frequency-resolved energy enhancement over random input wavefront, in a $10 \times 10$ \textmu m$^2$ target region centered at depth $z$ = 30 \textmu m in the disordered waveguide, for the maximum deposition eigenchannel with center frequency $\omega_0$ and bandwidth $\Delta \omega$. Both frequency detuning $|\omega - \omega_0|$ and the input bandwidth $\Delta \omega$ are normalized by the spectral correlation width $\delta\omega$ at this depth. The normalized bandwidths are $\Delta \omega / \delta \omega$ = 0 (black), 1.5 (red), 4.5 (green), 9.0 (blue). Experimental results are averaged over $\omega_0$ within the wavelength range of 1551 nm - 1556 nm and over two disorder realizations. }
	\end{figure}
	
	We consider a target area of dimension 10 \textmu m by  10 \textmu m centered at a depth of 30 \textmu m. The input wavelength is scanned from $\lambda$ =  1551 nm to 1556 nm with a step size of 0.1 nm. At each wavelength, we measure the deposition matrix. From the frequency-resolved matrix $\mathcal{Z(\omega)}$, we construct the broadband deposition matrix $\mathcal{A}$ for a given frequency interval of width $\Delta \omega$  centered at $\omega_0$, assuming constant intensity $I(\omega)$ within $\Delta \omega$. We compute the eigenvector of $\mathcal{A}$ with the largest eigenvalue corresponding to the maximal broadband deposition channel. Taking this eigenvector as the input wavefront and multiplying it by $\mathcal{Z(\omega)}$ gives the field distribution in the target region at frequency $\omega$. Spatial integration of the field intensity in the target provides the deposited energy as a function of frequency detuning $ |\omega - \omega_0|$.  
	Results are shown in Fig.~\ref{dep_spec} for different bandwidths. In the monochromatic case ($\Delta \omega \approx 0$), the energy deposition eigenchannel is highly sensitive to the frequency detuning, decaying quickly as $\omega$ deviates from $\omega_0$ (black curve). The full width at half maximum gives $\delta \omega \simeq$ 381 rad/ns. Both monochromatic frequency detuning from $\omega_0$ and input bandwidth $\Delta \omega$ are normalized by $\delta \omega$. In Fig.~\ref{dep_spec}, red, green, and blue curves represent the energy deposited by maximal eigenvectors of $\mathcal{A}$ with increasing bandwidth $\Delta \omega / \delta \omega$ = 1.5, 4.5, and 9. Within the designated frequency range, the energy enhancement remains nearly constant and well above 1. Even for $\Delta\omega / \delta\omega = 9$, the maximum eigenchannel of $\mathcal{A}$ sustains a 6-fold energy enhancement over random wavefront illuminations. Beyond $\Delta \omega$, the energy enhancement drops but still remains well above 1 for very large frequency detuning. These results reveal that the broadband deposition channels can sustain high energy enhancement over a large spectral range, greatly outperforming the monochromatic deposition channel across the majority of the input bandwidth $\Delta \omega$.
	
	\begin{figure}[b]
		\includegraphics[width=.48\textwidth]{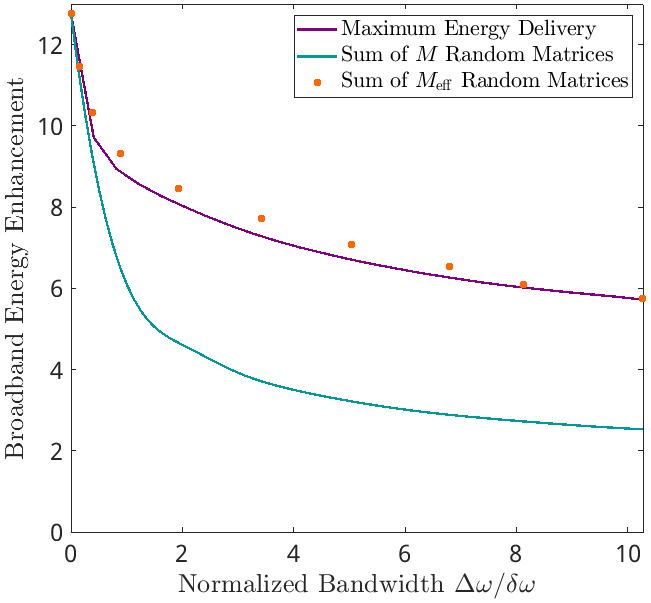}
		\caption{\label{max_dep} \textbf{Bandwidth dependence of targeted energy delivery.} Maximum enhancement of energy within input bandwidth $\Delta \omega$ normalized by $\delta \omega$, in a $10 \times 10$ \textmu m$^2$ target region centered at depth $z$ = 30 \textmu m in the disordered waveguide, is given by the ratio of the largest eigenvalue $\zeta_\text{max}$ of the experimentally measured broadband deposition matrix $\mathcal{A}$ to the mean eigenvalue $\langle \zeta \rangle$ (purple line). Approximating $\mathcal{A}$ by a sum of $M=\Delta\omega/\delta\omega+1$ random matrices underestimates the energy enhancement (blue line). Reducing $M$ to $M_{\text{eff}}$, to account for long-range spectral correlation, recovers the maximal eigenvalue of $\mathcal{A}$ (red symbols) and agrees to the experimental enhancement.}
	\end{figure}
	
	\section{Bandwidth Dependence}
	
	To further investigate the broadband energy deposition, we calculate the maximum eigenvalue $\zeta_{\text{max}}$ of $\mathcal{A}(\Delta \omega)$, which gives the largest energy that can be delivered to the target within the bandwidth of $\Delta \omega$. Since the mean eigenvalue $\langle \zeta \rangle$ corresponds to the target energy under random wavefront illuminations, $\zeta_{\text{max}} / \langle \zeta \rangle$ is equal to the broadband energy enhancement in the target. Figure~\ref{max_dep} shows that $\zeta_{\text{max}} / \langle \zeta \rangle$ decays with the normalized bandwidth $\Delta \omega / \delta \omega$ for a $10 \times 10$ \textmu m$^2$ target region centered at depth $z$ = 30 \textmu m in the disordered waveguide (purple curve). 
	
	To capture the impact of spectral correlations on the scaling of the energy enhancement with bandwidth, we model $\mathcal{A}$ in Eq.~\eqref{A} as a sum of $M_\text{eff}$ uncorrelated matrices: $\mathcal{A}\simeq \tilde{\mathcal{A}}$, where
 \begin{equation}
 \tilde{\mathcal{A}} = \sum_{m=1}^{M_\text{eff}}\tilde{\mathcal{Z}}^\dag_m\tilde{\mathcal{Z}}_m.
 \end{equation}
 Each random matrix $\tilde{\mathcal{Z}}_m$ represents an independent spectral channel, numerically generated with the same eigenvalue statistics as the measured $\mathcal{Z}(\omega)$ (for additional details see Appendix C). We plot the energy enhancement of the maximal eigenvector of $\tilde{\mathcal{A}}$ in Fig.~\ref{max_dep} for $M_\text{eff}=M = {\Delta\omega}/{\delta\omega} + 1$ (cyan curve). The result is notably smaller than the experimental one (purple curve). For a bandwidth of $\Delta\omega / \delta\omega = 10$, the actual energy enhancement is more than twice that calculated from the random matrix model.
	
Such dramatic discrepancy results from the lack of correlations between the $M$ random matrices $\tilde{\mathcal{Z}}_m$ used to generate $\tilde{\mathcal{A}}$. Correlations between matrices $\mathcal{Z}(\omega)$ in $\mathcal{A}$ originate from long-range spectral correlations of multiply-scattered waves in a diffusive system \cite{Stephen1987, Berkovits1994}. To account for this, we rescale the number of independent spectral channels $M$ within $\Delta \omega$ to $M_\text{eff}<M$~\cite{Popoff2014, Hsu2015}. 
 
 $M_\text{eff}$ is determined explicitly by the long-range spectral correlation function $C_2(z,|\omega_1-\omega_2|)$ at depth $z$~\cite{Hsu2015}, 
	\begin{equation}
	\frac{1}{M_\text{eff}}  =\frac{\bar{C}_2(z,\Delta\omega)}{\bar{C}_2(z,0)},
	\label{Meff_RMT}
	\end{equation}
 where $\bar{C}_2(z,\Delta\omega)=\iint_{\Delta\omega}d\omega_1d\omega_2C_2(z,|\omega_1-\omega_2|)/\Delta\omega^2$.
	In Appendix C, we present an analytic expression for $C_2(z,|\omega_1-\omega_2|)$, thereby providing an explicit equation to compute $M_\text{eff}$. 
	Using this expression without any free parameter, we obtain the enhancement plotted by circles in Fig.~\ref{max_dep} and recover the experimental result for broadband enhancement to a remarkable degree. Such agreement confirms the importance of long-range spectral correlations for delivering energy over large bandwidths. In particular, in the broadband limit $\Delta \omega \gg \delta \omega$, we find $M_\text{eff} \varpropto \sqrt{\Delta \omega/\delta \omega} \simeq \sqrt{M}$ (see Appendix C for details). The reduced number of independent channels ($M_\text{eff} \ll M$) enables a much greater improvement in the maximum energy deposition using a single wavefront than expected for $M$ uncorrelated channels. 
 
 In the next section, we discuss in more detail the link between the broadband long-range correlation function $\bar{C}_2(z,\Delta\omega)$ and the energy deposition enhancement, highlighting, in particular, their dependence on depth.
	
	\section{Depth Dependence of Energy Deposition}
	
	\begin{figure*}
		\includegraphics[width=\textwidth]{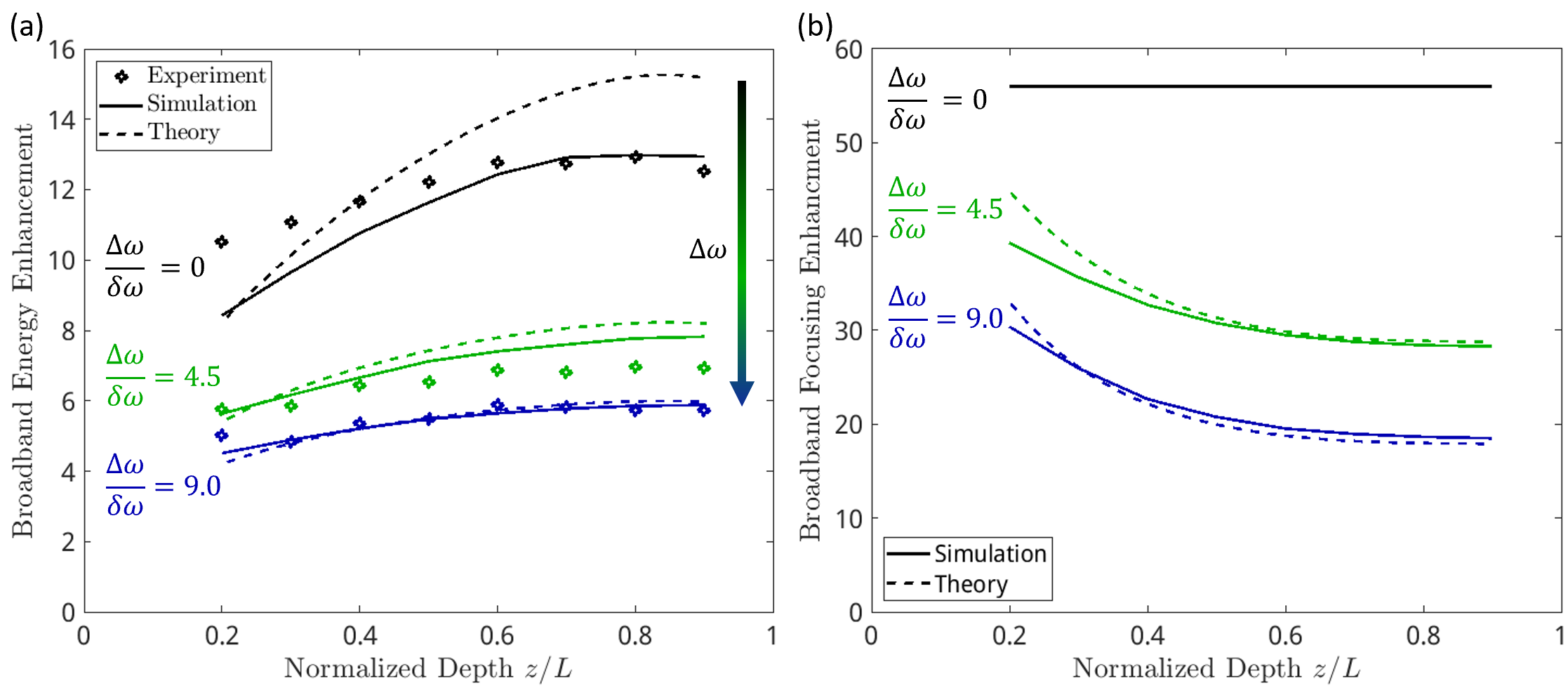}
		\caption{\textbf{Depth dependence of broadband energy delivery.} (a) Maximum enhancement of energy $\langle\zeta_\text{max}\rangle/\langle\zeta\rangle$, delivered to a $10 \times 10$ \textmu m$^2$ target region, as a function of depth $z$ (normalized by the length of disordered waveguide $L$). Input (normalized) bandwidth $ \Delta \omega / \delta \omega$ is fixed to 0 (black), 4.5 (green), 9.0 (blue). Experimental data (symbols) are compared to numerical results (solid line) and analytical prediction (dashed line). For $\Delta \omega / \delta \omega \gg 1$, broadband energy enhancement becomes nearly depth-independent. (b) Enhancement of broadband focusing to a wavelength-scale target as a function of normalized depth $z/L$, for fixed bandwidth $\Delta\omega/ \delta \omega$ = 0 (black), 4.5 (green), and 9.0 (blue). When $\Delta \omega / \delta \omega \ll 1$, focusing enhancement is independent of depth, and determined by the degree of control over the input wavefront which is equal to the number of waveguide modes $N$ = 56. As $\Delta \omega / \delta \omega$ increases, the focusing enhancement decays with the depth for $z<L/2$.  \label{Depth}}
	\end{figure*}
	
	The dependence of energy deposition on depth is of great importance for access deep inside scattering media. In the previous section, the depth of the target is fixed. Here we vary the depth $z$ while fixing the input bandwidth $\Delta \omega$. From the experimental data, we plot the maximum eigenvalue of  $\mathcal{A}$ versus $z$ for $\Delta\omega/ \delta \omega = 0,\, 4.5,\, 9$ in Fig.~\ref{Depth}a. Surprisingly, the energy enhancement increases with $z$, and the depth dependence nearly vanishes when $\Delta \omega \gg \delta \omega$. To confirm the experimental results, we conduct numerical simulations using KWANT, a Python package for wave transport simulations \cite{Groth2014} (for additional details see Appendix B). The simulation results, plotted by the solid lines in Fig.~\ref{Depth}a, have good agreement with experimental data (symbols), especially at large $\Delta\omega$. 
	
	To explain the depth dependence, we compute the broadband energy deposition analytically using filtered random matrix theory \cite{Goetschy2013, Hsu2015}. A complete derivation provided in Appendix C reveals that for extended-area deposition in the broadband limit, the energy enhancement depends only on $\bar{C}_2(z,\Delta\omega)$,
\begin{equation}
	\frac{\langle\zeta_{\text{max}}\rangle}{\langle\zeta\rangle} \simeq \gamma+3\left(\frac{\pi}{2}\right)^{2/3}\gamma^{1/3}-2,
	\end{equation}
 where $\gamma=3N\bar{C}_2(z,\Delta\omega)/2$.
	 The theoretical predictions, plotted by the dashed lines in Fig.~\ref{Depth}a, match the experimental and numerical results, predicting in particular that the depth dependence nearly vanishes at a large bandwidth. In the limit of $\Delta\omega/\delta\omega \gg 1$, we prove that $\gamma \simeq (16/\pi)L_{\Delta \omega}/\ell_t $, where the coherence length $L_{\Delta \omega}=\sqrt{D/2\Delta \omega} = \sqrt{D \, \tau_c /2}$ is equal to the diffusion distance that broadband light travels over the coherence time $\tau_c$. The diffusion coefficient is $D=\ell_t v/2$, where $v$ is the energy velocity.   As a result, $\langle\zeta_{\text{max}}\rangle/\langle\zeta\rangle$ becomes independent of the depth $z$. These results illustrate that long-range spatial and spectral correlations facilitate broadband energy delivery to extended targets.
	
	The depth dependence stands in sharp contrast to broadband focusing to a single speckle of wavelength scale.  Figure~\ref{Depth}b shows the focusing enhancement (ratio of maximum energy within the focal spot over the mean value), obtained from numerical simulations, for different bandwidths.  As $\Delta \omega$ is increased, the depth dependence becomes more pronounced. For $\Delta \omega / \delta \omega = 9$, the focusing enhancement decays rapidly with the normalized depth $z/L$. Such variation originates from the fact that the broadband focusing is determined by the short-range correlation function $C_1(z,|\omega_1-\omega_2|)$, which has a distinct depth dependence from $C_2(z,|\omega_1-\omega_2|)$. We compute the focusing enhancement analytically from the expression of $C_1(z,|\omega_1-\omega_2|)$ derived in Appendix C. In Fig.~\ref{Depth}b, the analytical results (dashed line) agree well with the numerical results (solid line). In particular, in the limit of thick samples ($L \gg L_{\Delta \omega}, z$), the focusing enhancement to a single speckle reduces to $4 N (L_{\Delta \omega}/z)^2$. These results indicate that when long-range spatial and spectral correlations are dominant, the benefits of single wavefront shaping for broadband energy delivery persist at any depth.

	\section{Effect of Dissipation}
	
	\begin{figure}
		\includegraphics[width=.48\textwidth]{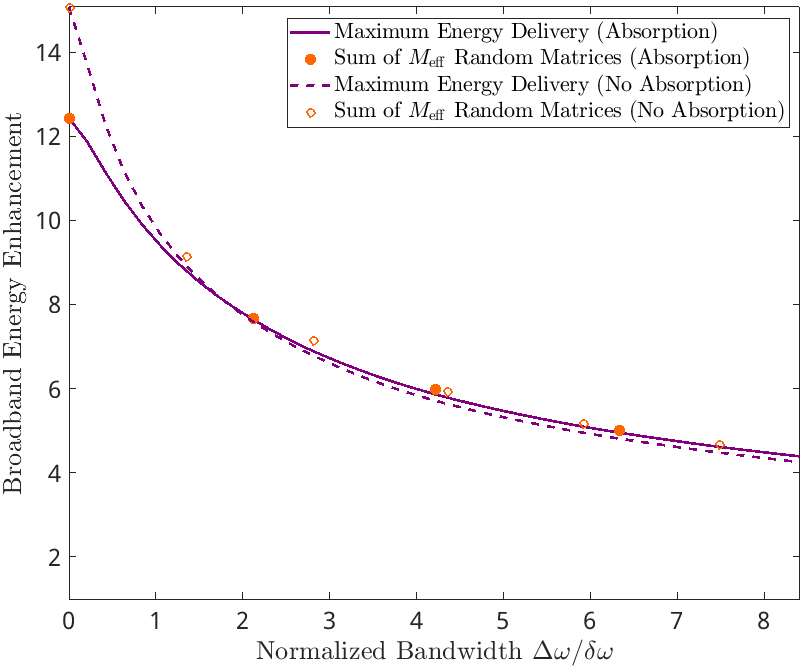}
		\caption{\textbf{Effect of dissipation on energy delivery.} Comparison of numerical results with (solid lines) and without (dashed lines) dissipation for maximum energy enhancement $\langle\zeta_{\text{max}}\rangle/\langle\zeta\rangle$,  in a $10 \times 10$ \textmu m$^2$ target region centered at the normalized depth $z/L = 0.6$. The diffusive dissipation length $\xi_a$ = 28 \textmu m is equal to the experimental value. For $\Delta \omega / \delta \omega \ll 1$, dissipation lowers the energy enhancement. However, for $\Delta \omega / \delta \omega \gg 1$, the energy enhancement becomes insensitive to dissipation. $M_\text{eff}$ successfully predicts the energy enhancement from the random matrix model both with (solid circles) and without (open circles) absorption. \label{Abs}}
	\end{figure}
	
	Finally, we investigate how dissipation affects broadband energy delivery. In our experiment, energy loss due to out-of-plane scattering can be described by a diffusive dissipation length $\xi_a = \sqrt{\ell_t \ell_a/2} = 28$ \textmu m, where $\ell_a$ is the ballistic dissipation length \cite{Yamilov2014}. Since $L \gtrsim\xi_a$, the dissipation is not negligible. Put differently, the dissipation time $\tau_a = \ell_a/v$ is comparable to the diffusion dwell time of light measured at the output of the waveguide, $\tau_d = L^2/\pi^2D$. To compare a system with dissipation to one without dissipation, we conduct numerical simulations where dissipation can be switched off. 
	
	In Fig.~\ref{Abs}, we plot the numerical results of energy enhancement with and without dissipation as a function of input bandwidth $\Delta \omega$ for a $10 \times 10$ \textmu m$^2$ target at depth $z/L = 0.6$. The dissipation length of $\xi_a = 28$ \textmu m in the simulations is close to the experimental value. For monochromatic light, dissipation lowers the energy enhancement, in accordance with the previously demonstrated reduction of transmission enhancement in the presence of weak and moderate dissipation  \cite{Yamilov2016}. As $\Delta \omega$ is increased, however, the broadband energy enhancement with dissipation approaches that without dissipation. For confirmation, we calculate the energy enhancement from random matrices by rescaling the number of independent spectral channels by $M_\text{eff}$. Regardless of the absorption, $M_\text{eff}$ calculated from $C_2$ gives the energy enhancement in good agreement with the numerical results, as shown in Fig.~\ref{Abs}.
 
    The convergence of enhancement with and without absorption at large bandwidths indicates that dissipation has little impact on the maximum energy delivery achievable by wavefront shaping. We find that this is because the long-range correlation $C_2$ loses its dependence on $\xi_a$ in the broadband limit. A physical explanation is given in terms of the coherence time of light $\tau_c = 1/\Delta \omega$. For narrow-band input, $\tau_c \gg \tau_a$, dissipation weakens the interference effect responsible for $C_2$ by attenuating the scattered waves of long paths. More precisely, it reduces the probability of two diffusive paths crossing inside the sample, an effect at the origin of $C_2$ correlations.   The energy enhancement induced by this constructive interference is, therefore reduced. As $\Delta \omega$ is increased, $\tau_c$ becomes shorter. Once $\tau_c < \tau_a$, the interference effect is limited by decoherence rather than dissipation: the exchange of field partners occurs inside a volume of thickness $L_{\Delta \omega} < \xi_a$, and the $C_2$ contribution saturates to $\bar{C}_2(z,\Delta\omega)\simeq (32/3\pi N) L_{\Delta \omega}/\ell_t$. Consequently, the dependence of energy enhancement on dissipation vanishes. This result also contrasts sharply with what is obtained for broadband focusing, which preserves a non-negligible absorption dependence even in the regime $\tau_c < \tau_a$ (see Appendix C).

	\section{Discussion and Conclusion}
	
	We have introduced the broadband deposition matrix $\mathcal{A}$, which identifies a single wavefront that delivers the maximum energy from a broadband source deep into a diffusive system. For a target of size much larger than the wavelength, long-range spatial and spectral correlations enhance the energy delivery. Both the dependence on target depth and on (uniform) dissipation in the sample vanish in the broadband (fast decoherence) limit. These results contrast those for broadband focusing to a wavelength-scale target, which is dictated by short-range correlations, leading to a quick decay of focusing enhancement with depth. This difference highlights that contributions of long-range correlations to broadband energy deposition will dominate at large depths, a result that is not necessarily true for monochromatic energy delivery.

	While this study is performed on two-dimensional diffusive systems, the general conclusion and theoretical model are applicable to three-dimensional (3D) systems. Although it is very difficult to measure the deposition matrix inside a 3D system \cite{hong2018}, our theoretical model can still predict the fundamental limit for broadband energy delivery to a large target located at any depth, proportional to $N\bar{C}_2(z,\Delta\omega)$ at leading order, thereby providing an upper bound with which to constrain wavefront optimization. Note that for open slab geometries in realistic applications, incomplete channel control must be accounted for in the evaluation of $\bar{C}_2(z,\Delta\omega)$~\cite{Popoff2014, Hsu2015}. In any case, the maximum eigenvalue of the partial broadband deposition matrix still sets the highest possible energy delivered to a target, and the corresponding eigenvector provides the input wavefront. Our approach may also be applied to systems with tailored disorder or scattering, such as photonic crystals \cite{Uppu2021}. Finally, our methods and results apply more generally to other types of waves, including acoustic waves, microwaves, and electron waves, which possess long-range spatial and temporal correlations. Overcoming the hurdle of fast temporal decoherence of these waves would have implications for practical applications, including deep-tissue imaging \cite{Yu2015, Yoon2020}, optogenetic control of neurons \cite{2011_Fenno, Yoon2015, 2017_Pegard_NC, Ruan2017}, laser microsurgery \cite{2004_Yanik, Yu2015}, and photothermal therapy \cite{Pernot2007} deep inside complex media.
	
    \section*{Acknowledgements}

    We would like to thank Chun-Wei Chen, Kyungduk Kim, SeungYun Han, Liam Shaughnessy, Peng Miao, Sam Halladay, and Zehua Lai for valuable discussions. This work is supported partly by the National Science Foundation (NSF) under Grant No. DMR-1905465 and by the Office of Naval Research (ONR) under Grant No. N00014-221-1-2026.
    
    \appendix

    \section{Experimental Measurements}

    \subsection{Sample Fabrication}
    
    The waveguiding structures used in our experiment are fabricated on a silicon-on-insulator wafer using a combination of electron beam lithography and reactive ion etching \cite{Bender2022_2, Bender2022}. Each structure consists of four sections: a ridge waveguide with width of 300 \textmu m and length of 15 mm, an adiabatic-taper ($15^{\circ}$ angle) where the waveguide width gradually decreases from 300 \textmu m to 15 \textmu m, a weak-scattering buffer region with width 15 \textmu m and length 25 \textmu m, and a strong-scattering region with width 15 \textmu m and length 50 \textmu m. last two segments have randomly distributed air holes of diameter 100 nm, and air filling fractions of $0.55\%$ and $5.5\%$, respectively. Both regions support 56 spatial modes around the wavelength $\lambda$ = 1550 nm. Light is confined by a photonic-crystal triangle-lattice boundary in the last three sections of the structure. The photonic-crystal boundary consists of 16 layers of air holes with a radius of 155 nm and a lattice constant of 440 nm. The boundary provides a complete 2D photonic bandgap for TE polarized light of wavelength 1120 nm through 1580 nm. The tapered and buffer regions facilitate mode mixing and excitation of high-order modes in the waveguide \cite{Sarma2016}. The transport mean free path $l_t$ = 33 \textmu m in the buffer region, and $l_t$ = 3.3 \textmu m in the strong-scattering region \cite{Bender2022_2, Bender2022}.  
    
    \begin{figure*}
    	\includegraphics[width=.96\textwidth]{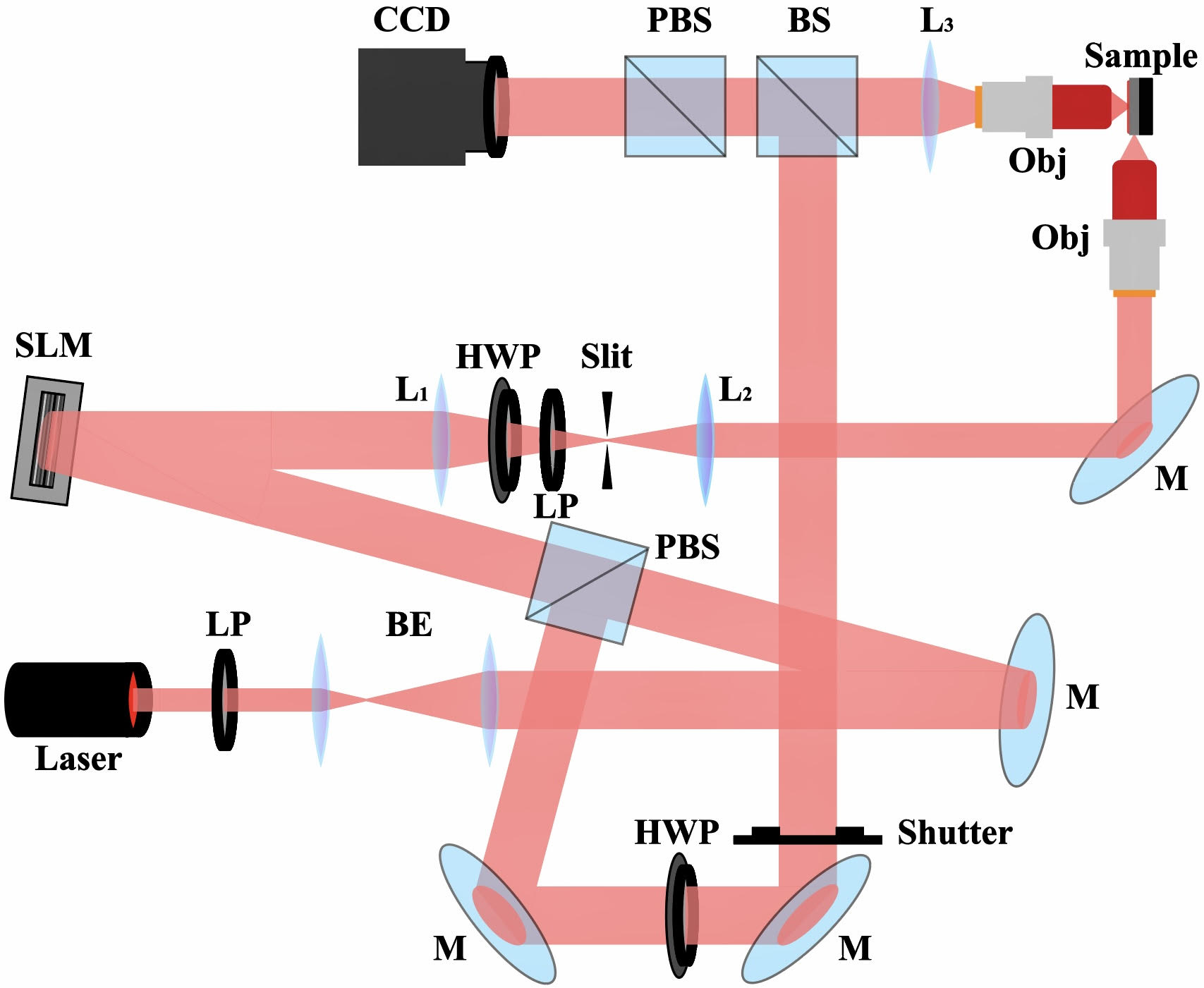}
    	\caption{\textbf{Experimental Setup.} A monochromatic, linearly-polarized beam from a tunable laser is split into two by a polarizing beam splitter (PBS). One beam is modulated by a spatial light modulator (SLM), then its polarization direction is rotated by a half-wave-plate (HWP) before being launched to a planar waveguide in a silicon-on-insulator wafer through the side. The other beam is used as a reference beam, and it is combined with the out-of-plane scattered light from the waveguide by an unpolarized beam splitter (BS). After passing through a second PBS, the linearly-polarized interference pattern is recorded by a CCD camera. By varying the relative phase between the reference and the signal with the SLM, the field distribution inside the 2D disordered waveguide is recovered from the interference patterns, giving the field deposition matrix at a single frequency. By scanning the input laser frequency, the broadband deposition matrix is built from frequency-resolved deposition matrix. LP: linear polarizer; BE: beam expander; M: mirror; Obj: long-working-distance objective lens (Mitutoyo M Plan APO NIR HR100x, NA = 0.7); L: lens with focal length = 400 mm ($L_1$), 75 mm ($L_2$), and 100 mm  ($L_3$).  \label{setup}}
    \end{figure*}
    
    \subsection{Optical Setup}
    
    Figure \ref{setup} is a detailed schematic of our experimental setup, which has been described in Refs.~\cite{Bender2022_2, Bender2022}. The monochromatic light from a
    wavelength-tunable laser (Keysight 81960A) is split into two beams. One beam is modulated by a phase-only spatial light modulator (Hamamatsu LCoS X 10468) and then injected into one of the waveguides via the edge of the wafer. The other beam is used as a reference beam which is spatially overlapped with the out-of-plane scattered light from the diffusive waveguide. A CCD camera (Allied Vision Goldeye G-032 Cool) records the resulting interference pattern, from which the complex field profile across the diffusive waveguide is obtained. 
    
    \subsection{Deposition Matrix Measurement}
    
    By sequentially applying an orthogonal set of one-dimensional (1D) phase patterns to the 128 SLM macropixels, and measuring the field within the sample, we acquire a matrix that maps the field from the SLM to the field inside the disordered waveguide $\mathcal{Z}_{\rm slm \rightarrow \rm int}$. This operator encompasses information about both the light transport inside the waveguide and the light propagation from the SLM to the waveguide. To separate these, we add an auxiliary weakly-scattering region in front of the diffusive region called the `buffer' region. We recover the field right in front of the strongly-scattering region by imaging the light scattered out-of-plane from the buffer. The length of the buffer region is 25 \textmu m, which is shorter than its 32 \textmu m-length transport mean free path. Therefore, light only experiences single scattering in the buffer, so the diffusive wave transport in the original disordered region is not appreciably altered.
    
    With access to the field inside the buffer, we can construct the matrix relating the field on the SLM to the buffer, $\mathcal{Z}_{\rm slm \rightarrow \rm buff}$. From $\mathcal{Z}_{\rm slm \rightarrow \rm int}$, we can also construct the deposition matrix to a selected target region, $\mathcal{Z}_{\rm slm \rightarrow \rm tar}$, which maps the field from the SLM to a region near the end of the diffusive waveguide. With these we calculate the matrix that maps the field from the buffer to the target, $\mathcal{Z}_{\rm buff\rightarrow tar}= \mathcal{Z}_{\rm slm\rightarrow tar} \, \mathcal{Z}_{\rm slm\rightarrow buff}^{-1}$, using Moore-Penrose matrix inversion. We keep only the top N = 56 singular vectors with large singular values ($N$ is the number of waveguide modes), to suppress the experimental noise that lies primarily in the singular vectors with small singular values. This procedure is repeated for 50 wavelengths in the $1551$-$1556$ nm range with a step size of $0.1$ nm, to obtain the frequency-resolved deposition matrix $\mathcal{Z}_{\rm buff\rightarrow tar}(\omega)$.
    
    In the sequential measurement of $\mathcal{Z}_{\rm buff\rightarrow tar}(\omega)$, the $N = 56$ dimensional subspace kept in the matrix inversion will not be the same for all measured frequencies due both to experimental noise and waveguide dispersion, making it very difficult to construct the broadband deposition matrix $\mathcal{A}_{\rm buff\rightarrow tar}(\Delta \omega)$. To obtain a common subspace of all frequencies, we calculate the extrinsic mean for vector subspaces defined as the average of the projection matrices for each subspace \cite{978385}.  This is valid as long as dispersion is relatively weak inside the buffer. When we perform the Moore-Penrose pseudo inverse for the deposition matrix for each frequency, we project onto the extrinsic subspace mean so that all deposition matrices share the same N-dimensional space in the buffer region. This procedure allows us to not only build the broadband deposition matrix but also to compute the field distribution at frequency $\omega_2$ for the input wavefront given by the deposition eigenchannel at $\omega_1$.
    
    The average intensity in the target region is recorded during the experiment before matrix inversion. It is integrated spatially and spectrally to give the broadband energy within the target region under random wavefront illumination.

    \section{Numerical Simulations}
    
    We perform numerical simulations of wave propagation in planar waveguides using KWANT \cite{Groth2014, Yamilov2014}. 
    Simulated waveguides have the same dimensions and parameters as in the experiment: width $W$, length $L$, refractive index, and number of waveguide modes $N$. The transport mean free path $l_t$ = 3.3 \textmu m, and the diffusive dissipation length $\xi_a$ = 28 \textmu m. In the numerical simulations, we can turn off the dissipation by setting $\xi_a = \infty$. To be consistent with our experiment, the deposition matrices map the fields from a buffer region to 10 \textmu m by 10 \textmu m target regions at varying depths. All numerical results are averaged over 200 disorder realizations.

    \section{Theoretical predictions}
    
    \subsection{Filtered random matrix (FRM) model}
    
    Following the approach developed in Ref.~\cite{Hsu2015} for the broadband transmission matrix, we model the broadband deposition matrix
    \be
    \mathcal{A} = \int_{\Delta\omega} \frac{d\omega}{\Delta \omega}\, \mathcal{Z}^\dag(\omega)\mathcal{Z}(\omega)
    \label{EqMatA}
    \ee
    as a sum of $M_{\text{eff}}$ uncorrelated matrices $\mathcal{Z}^\dag_m\mathcal{Z}_m$ statistically equivalent to $\mathcal{Z}^\dag\mathcal{Z}$,
    \be
    \mathcal{A} \simeq \sum_{m=1}^{M_\text{eff}}\mathcal{Z}^\dag_m\mathcal{Z}_m. 
    \ee
    It follows that the Stieltjes transform $g_{\mathcal{A}}(w)$ of the eigenvalue distribution of $\mathcal{A}$, $p_{\mathcal{A}}(\zeta) = -\lim_{\epsilon \to 0^+}\text{Im}\left[g_{\mathcal{A}}(\zeta + i\epsilon)/\pi\right]$,  obeys the following implicit equation~\cite{Hsu2015}
    \be
    g_{\mathcal{A}}(w)=M_\text{eff} \, , g_{\mathcal{Z}^{\dag}\mathcal{Z}}\left(w + \frac{M_\text{eff}-1}{g_{\mathcal{A}}(w)}\right),
    \label{EqResolvent1}
    \ee
    where $g_{\mathcal{Z}^{\dag}\mathcal{Z}}(w)$ is the Stieltjes transform of the matrix $\mathcal{Z}^\dag\mathcal{Z}$.
    
    Equation~\eqref{EqResolvent1} implies in particular that the variances of the eigenvalues of $\mathcal{A}$ and $\mathcal{Z}^{\dag}\mathcal{Z}$, noted  $\zeta$ and $\zeta_0$ respectively, are related through the  relation~\cite{Popoff2014}
    \be
    \text{Var}(\zeta)=\frac{\text{Var}(\zeta_0)}{M_\text{eff}}.
    \label{EqVar}
    \ee 
    As both $\text{Var}(\zeta)$ and $\text{Var}(\zeta_0)$ can be computed microscopically using a Feynman path-type decomposition of the deposition matrix $\mathcal{Z}(\omega)$, we can use Eq.~\eqref{EqVar} to find an explicit expression of $M_\text{eff}$. The latter is given by~\cite{Popoff2014, Hsu2015}
    \be
    \frac{1}{M_\text{eff}}=\iint_{\Delta\omega}\frac{d\omega_1d\omega_2}{\Delta\omega^2}\frac{C^{(T)}(z,\omega_1,\omega_2)}{C^{(T)}(z,\omega_0,\omega_0)},
    \label{EqMeffExact}
    \ee
    where 
    \be
    C^{(T)}(z,\omega_1,\omega_2)=\frac{\left<I(z,\omega_1) I(z,\omega_2) \right>}{\left<I(z,\omega_1) \right>\left< I(z,\omega_2) \right>} -1
    \label{EqTotalCorrelation}
    \ee
     is the spectral correlation of the total intensity  $I(z,\omega)=\sum_b \vert\mathcal{Z}_{ba}(\omega)\vert^2$ deposited at depth $z$,  with $\left< \dots \right>$ representing the average over different realizations of the disorder. The correlation function $C^{(T)}$ contains both short- and long-range contributions, but if the number of waveguide channels is large enough ($NC_2 \gg1$),  $C^{(T)}$ is dominated by the long-range component noted $C_2$ and evaluated explicitly in the next section. Then, Eq.~\eqref{EqMeffExact} reduces to
    \be
    \frac{1}{M_\text{eff}}\simeq \frac{\bar{C}_2(z,\Delta\omega)}{\bar{C}_2(z,0)},
    \label{EqMeff}
    \ee 
     where $\bar{C}_2(z,\Delta\omega)=\iint_{\Delta\omega}d\omega_1d\omega_2C_2(z,|\omega_1-\omega_2|)/\Delta\omega^2$.
     
     Combining Eqs.~\eqref{EqResolvent1} and ~\eqref{EqMeff} gives access to the eigenvalue distribution $p_{\mathcal{A}}(\zeta)$ once $g_{\mathcal{Z}^{\dag}\mathcal{Z}}(w)$ is known. It has been shown in Ref.~\cite{Bender2022} that the latter is itself the solution of the implicit equation
     \begin{widetext}
    \be
    g_{\mathcal{Z}^{\dag}\mathcal{Z}}(w)=
    \frac{w\,m\,g_{\mathcal{Z}^{\dag}\mathcal{Z}}(w)+1-m}{w\,m^2\,g_{\mathcal{Z}^{\dag}\mathcal{Z}}(w)}g_{t_0^\dagger t_0}
    \left[
    \frac{\left[w\,m\,g_{\mathcal{Z}^{\dag}\mathcal{Z}}(w)+1-m\right]^2}{w\, m^2\,g_{\mathcal{Z}^{\dag}\mathcal{Z}}(w)^2}
    \right],
    \label{EqResolvent2}
    \ee 
     \end{widetext}
    where $t_0$ represents the transmission matrix of a virtual opaque disordered medium. Putting together Eqs.~\eqref{EqResolvent1} and~\eqref{EqResolvent2}, we find that $g_{\mathcal{A}}(w)$ obeys the FRM equation~\cite{Goetschy2013}
    \be
    \frac{N(w)}{D(w)}g_{t_0^\dagger t_0}
    \left[\frac{N(w)^2}{D(w)^2}
    \right]=1,
    \label{EqResolvent3}
    \ee
    where $N(w)=wm_1g_{\mathcal{A}}(w)+1-m_1$ and $D(w)=m_1g_{\mathcal{A}}(w)\left[wm_1g_{\mathcal{A}}(w)+m_2-m_1\right]$. The effective filtering parameters of this FRM model are $m_1=m/M_{\text{eff}}$ and $m_2=m$, where  $M_{\text{eff}}$ is given by Eq.~\eqref{EqMeff} and $m$ is~\cite{Bender2022}
    \begin{align}
    m&=\frac{\left< \zeta_0(L)\right>}{\left< \zeta_0(z)\right>}
    \nonumber
    \\
    &\simeq \frac{z_0/\xi_a}{\text{sinh}[(L-z)/\xi_a]+(z_0/\xi_a)\text{cosh}[(L-z)/\xi_a]},
    \end{align}
    where $z_0=\pi \ell_t/4$ is the extrapolation length and $\xi_a=\sqrt{\ell_t\ell_a/2}$ the diffusive dissipation length. In addition, the Stieltjes transform of the virtual opaque medium appearing in Eq.~\eqref{EqResolvent3} is 
     \be
    g_{t_0^\dagger t_0}(w)=\frac{1}{w}-\frac{\bar{\tau}_0}{w\sqrt{1-w}}\textrm{Arctanh}\left[
    \frac{\textrm{Tanh}(1/\bar{\tau}_0)}{\sqrt{1-w}}\right],
     \label{ResolventTM}
    \ee
    where  $\bar{\tau}_0$ is given by
    \be
    \bar{\tau}_0=\frac{2m}{3(NC_2(z,0)+2m)}.
    \label{EqTau0}
    \ee
    This expression guarantees that the normalized variance $\text{Var}(\zeta_0)/\left< \zeta_0 \right>^2$ is equal to $1 + NC_2(z,0)$~\cite{Bender2022}.
    
    In full generality, the eigenvalue distribution $p_{\mathcal{A}}(\zeta)$ follows from the numerical solution of Eq.~\eqref{EqResolvent3}. An approximate analytical solution can be found in the regime $\ell_t \ll L-z$ where $m\ll 1$. In that case, it was shown in Refs.~\cite{Bender2022, Bender2022_2} that $p_{\mathcal{A}}(\zeta)$ depends  on the parameter $\gamma=m_1/\bar{\tau_0}$ only. According to Eqs.~\eqref{EqMeff} and ~\eqref{EqTau0}, this parameter can be expressed as  $\gamma= m/M_{\text{eff}}\bar{\tau_0}\simeq 3NC_2(z,0)/2M_{\text{eff}} \simeq 3N\bar{C}_2(z,\Delta \omega)/2$. In particular, for $\gamma \gg1$, the upper edge of the distribution $p_{\mathcal{A}}(\zeta)$ can be expanded as~\cite{Bender2022}
    \be
    \frac{\langle\zeta_{\text{max}}\rangle}{\langle\zeta\rangle} \simeq \gamma+3\left(\frac{\pi}{2}\right)^{2/3}\gamma^{1/3}-2 + \mathcal{O} \left(\gamma^{-1/3} \right).
    \label{EqEnhancement}
    \ee
    We used this simplified expression together with the analytical expression of $\bar{C}_2(z,\Delta \omega)$ provided in the next section to generate the theoretical predictions shown in Fig.~4 of the main text. The expression of $\bar{C}_2(z,\Delta \omega)$ shows in particular that the condition  $\gamma \gg1$ used to derive Eq.~\eqref{EqEnhancement} is satisfied as long as $L_{\Delta \omega}\gtrsim\ell_t$, where  $L_{\Delta \omega}=\sqrt{D/2\Delta \omega}$ is the coherence length of the diffusive broadband light.
    
\begin{figure*}
    \includegraphics[width=\textwidth]{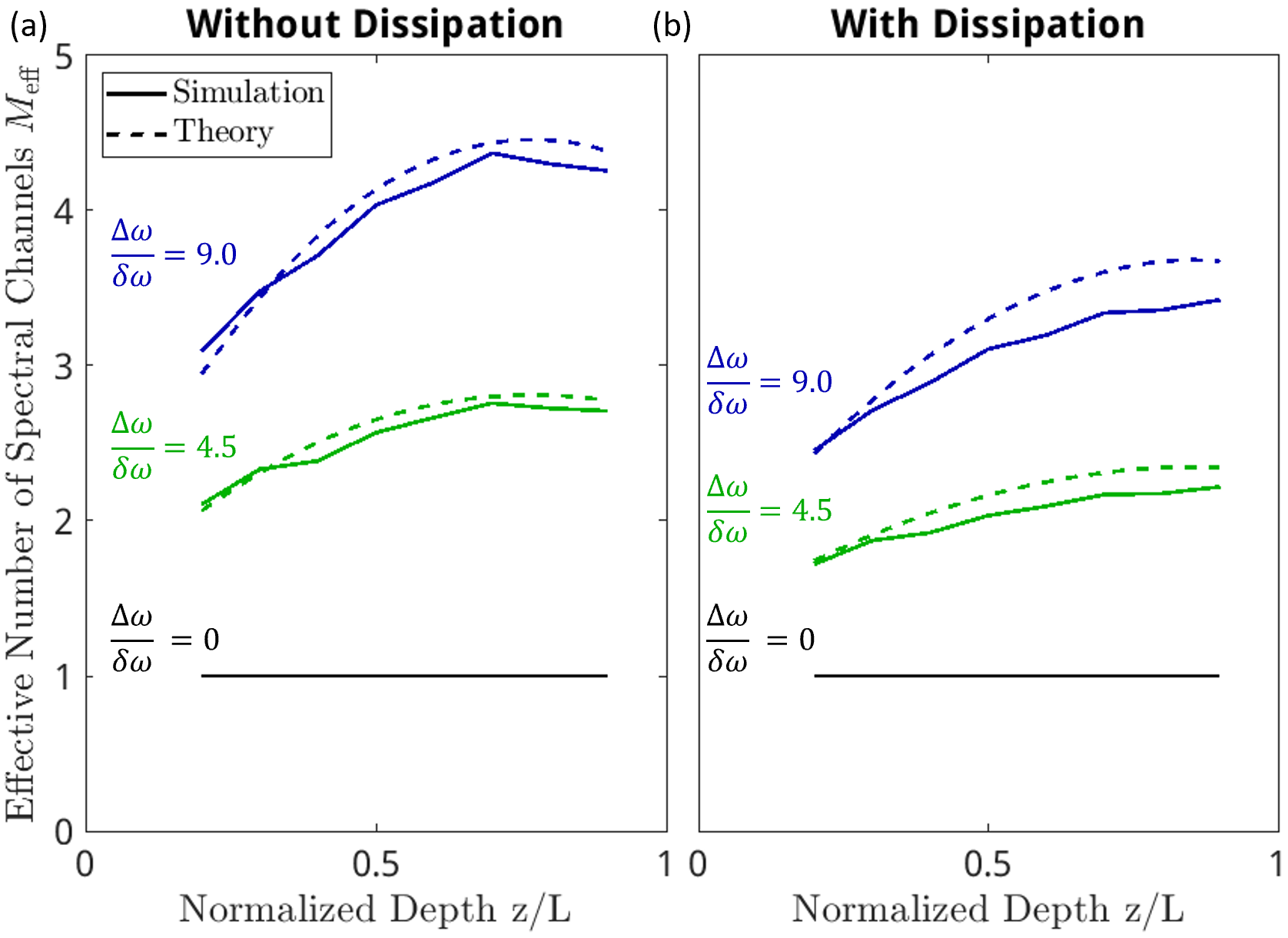}
    \caption{\textbf{Effective number of independent spectral channels $M_\text{eff}$.} Numerically simulated (solid line) and analytically predicted (dashed line) $M_\text{eff}$ as a function of normalized depth $z/L$ without dissipation (a) and with absorption (b). Input bandwidth $\Delta\omega / \delta\omega$ = 0 (black), 4.5 (green), and 9.0 (purple), where $\delta\omega$ is the spectral correlation width at $z/L = 3/5$. Depth dependence of $M_\text{eff}$ is stronger at larger bandwidth. For fixed $\Delta \omega/\Delta \omega$, $M_\text{eff}$ with dissipation is smaller than that without dissipation. \label{Abs_Meff}}
    \end{figure*}

    The quantity $M_\text{eff}$ characterizes the contribution of long-range spectral correlations on the maximum broadband deposition. Figure \ref{Abs_Meff} plots $M_\text{eff}$ as a function of $z/L$ for varying $\Delta \omega$. Both with and without dissipation, theoretical predictions given by Eqs.~\eqref{EqMeff} and ~\eqref{EqC2Abs} agree with numerical simulation results. For monochromatic light ($\Delta \omega = 0$), $M_{\rm eff}=1$ at any depth, regardless of dissipation. Once $\Delta \omega/ \delta \omega > 1$, $M_{\rm eff}$ increases with depth. In the absence of dissipation, $M_{\rm eff}$ reaches the maximum at $z/L \simeq 0.7$. In a dissipative system, $M_\text{eff}$ is smaller for the same $\Delta \omega /\delta\omega$, because the monochromatic component $\bar{C}_2(z,0)$ is reduced by absorption, while the broadband component $\bar{C}_2(z,\Delta\omega)$ depends only weakly on absorption (see next section for details). $M_\text{eff}$ increases monotonically with $z$ and reaches the maximum at $z \simeq L$.  At large bandwidth, the depth dependence of $M_\text{eff}$ gets stronger with or without absorption, even though the broadband energy enhancement becomes nearly depth invariant. This is explained by recognizing the strict dependence of $M_\text{eff}$ on $C_2(z,0)$ which has strong depth dependence, whereas $C_2(z,\Omega)$ is independent of depth [see  Eq.~\eqref{EqC2Approx}]. In particular, in the broadband limit and without absorption, we find
    \be
    M_\text{eff} \simeq \frac{3}{8}\frac{z}{L_{\Delta \omega}}
    \left(
    1-\frac{2}{3}\frac{z}{L}
    \right).
    \ee
    
    \subsection{Broadband long-range correlations inside the scattering sample}
    
     \begin{figure*}[t]
    \includegraphics[width=1\textwidth]{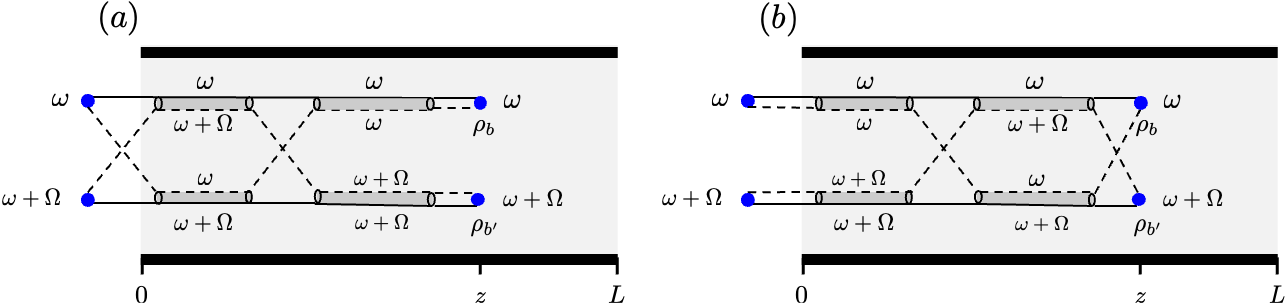}
    \caption{ {\bf Two contributions to long-range correlation.} Diagrams representing the long-range components $C_2^{(1)}(z,\Omega)$ in (a) and $C_2^{(2)}(z,\Omega)$ in (b) of  the four-field correlation function~\eqref{EqTotalCorrelation}.  Solid and dashed lines respectively represent the fields and their complex conjugates, shaded tubes represent diffusive paths and open circles represent scatterers. The exchange of field partners between diffusive paths corresponds to a Hikami box. Integration over the transverse coordinates $\rho_b$ and $\rho_{b'}$ of the deposition points located at depth $z$ must be performed to calculate the spectral correlation of the total intensity~\eqref{EqTotalCorrelation}.
    \label{FigDiagrams}
    }
    \end{figure*}
    
    The long-range component $C_2(z, \Omega)$ of the intensity correlator $C^{(T)}(z, \omega, \omega+ \Omega)$ defined in Eq.~\eqref{EqTotalCorrelation} contains two terms, $C_2^{(1)}(z, \Omega)$ and $C_2^{(2)}(z, \Omega)$, represented diagrammatically in Fig.~\ref{FigDiagrams}. In a waveguide geometry supporting $N$ transverse propagating modes, their explicit expressions read
     \begin{widetext}
    \begin{align}
    C_2^{(1)}(z,\Omega) = \frac{2}{gL\langle \mathcal{I}(z,0)\rangle^2}\int_0^Ldz'\vert \langle \mathcal{I}(z',\Omega)\rangle
    \vert^2\left[\partial_{z'}K(z,z',0)\right]^2,
    \label{EqC21}
    \\
    C_2^{(2)}(z,\Omega) = \frac{2}{gNL\langle \mathcal{I}(z,0)\rangle^2}\int_0^Ldz' \langle \mathcal{I}(z',0)\rangle
    ^2\vert\partial_{z'}K(z,z',\Omega)\vert^2,
    \label{EqC22}
    \end{align}
    \end{widetext}
    where $\langle \mathcal{I}(z,\Omega)\rangle =\left<E(z,\omega)E(z,\omega+\Omega)^* \right>$ is the field correlation at depth $z$. The latter can be expressed as $\langle \mathcal{I}(z,\Omega)\rangle=\int_0^Ldz'e^{-z'/\ell_t}K(z,z',\Omega)$, where $K(z,z',\Omega)$ is the Green's function of the diffusion equation
    \be
    \left(-\partial^2_z+\frac{1}{\xi^2_a}\right)-\frac{i\Omega}{D}K(z,z',\Omega) = \delta(z-z'),
    \ee
    with boundary conditions $\partial_z K(0,z',\Omega) = K(0,z',\Omega)/z_0$ and  $\partial_z K(L,z',\Omega) = -K(L,z',\Omega)/z_0$. The prefactors in Eqs.~\eqref{EqC21} and~\eqref{EqC22} involve the bare conductance $g=N\left<\tau\right>$, with the average transmittance (mean transmission eigenvalue) $\left<\tau\right>\simeq (\pi/2) \ell_t/L$. Note also the presence of the additional factor $1/N$ in $C_2^{(2)}(z,\Omega)$, which comes from the additional field crossing at depth $z$ shown in Fig.~\ref{FigDiagrams}(b); the latter forces the two integration points $\rho_b$ and $\rho_{b'}$ to be located in the same speckle grain, a condition that is absent for the diagram shown in Fig.~\ref{FigDiagrams}(a).

    In the diffusive limit $L\gg\ell_t$, integration in Eqs.~\eqref{EqC21} and~\eqref{EqC22} can be performed explicitly. In the absence of dissipation, we find
     \begin{widetext}
    \begin{align}
    C_2^{(1)}(z,\Omega) &=\frac{2}{g}\frac{1}{\tilde{L}}
    \frac{
    1
    }{\text{cosh}(\tilde{L})-\text{cos}(\tilde{L})}
    \left[
    \text{sinh}(\tilde{L} )-\text{sin}(\tilde{L})-\frac{L(L-2z)}{(L-z)^2}\left[ \text{sinh}(\tilde{L}-\tilde{z})-\text{sin}(\tilde{L}-\tilde{z} )\right]\right],
    \\
    C_2^{(2)}(z,\Omega) &=\frac{2}{g}\frac{1}{\tilde{L}}
    \frac{
    1
    }{\text{cosh}(\tilde{L})-\text{cos}(\tilde{L})} \frac{A(z,\Omega)+B(z,\Omega)}{N},
    \end{align}
    where $\tilde{L}=L/L_\Omega$, $\tilde{z}=z/L_\Omega$, with $L_\Omega=\sqrt{D/2\Omega}$, and 
    \begin{align}
    A(z,  \Omega)&=\left[
    \text{sinh}(\tilde{z})-\text{sin}(\tilde{z})+\frac{1}{2}(\tilde{L}-\tilde{z})^2
    \left[\text{sinh}(\tilde{z})+\text{sin}(\tilde{z})\right]\right]
    \frac{\text{cosh}(\tilde{L}-\tilde{z})-\text{cos}(\tilde{L}-\tilde{z})}{(\tilde{L}-\tilde{z})^2},
    \\
    B(z,  \Omega)&=\left[
    \text{sinh}(\tilde{L}-\tilde{z})-\text{sin}(\tilde{L}-\tilde{z})+\frac{1}{2}(\tilde{L}-\tilde{z})^2\left[\text{sinh}(\tilde{L}-\tilde{z})+\text{sin}(\tilde{L}-\tilde{z})\right]\right]
    \frac{\text{cosh}(\tilde{z})-\text{cos}(\tilde{z})}{(\tilde{L}-\tilde{z})^2}.
    \end{align}
    \end{widetext}
    The two contributions verify the equality $C_2^{(2)}(L,\Omega)=C_2^{(1)}(L,\Omega)/N$. Although this relation does not hold for $z\neq L$, we still have $C_2^{(2)}(z,\Omega) \ll C_2^{(1)}(z,\Omega)$ for $N\gg1$. As a result, we will neglect $C_2^{(2)}(z,\Omega)$ in the following and discuss the expression of $C_2(z,\Omega) \simeq C_2^{(1)}(z,\Omega)$ only. In the presence of dissipation, we find
     \begin{widetext}
    \be
    C_2^{(1)}(z,\Omega)=\frac{2}{g} 
    \frac{1}{\tilde{L}}
     \frac{1}{\text{cosh}(\alpha \tilde{L})-\text{cos}(\beta\tilde{L})}
     \left[I_1(z,\Omega)+\frac{\text{sinh}(z/\xi_a)^2}{\text{sinh}[(L-z)/\xi_a]^2}I_2(z,\Omega)\right],
     \label{EqC2Abs}
    \ee
    where $\alpha =\left[\left(1+4/\tilde{\xi_a}^4\right)^{1/2}+2/\tilde{\xi_a}^2\right]^{1/2}$, $\beta =\left[\left(1+4/\tilde{\xi_a}^4\right)^{1/2}-2/\tilde{\xi_a}^2\right]^{1/2}$, with $\tilde{\xi_a}=\xi_a/L_\Omega$, and
    \begin{align}
    I_1(z,\Omega)=&
     \frac{[-2+(\alpha \tilde{\xi}_a)^2] \text{sinh}(\alpha \tilde{L})
    +[2-(\alpha \tilde{\xi}_a)^2\text{cosh}(z/\xi_a)^2]
    \text{sinh}[\alpha(\tilde{L}-\tilde{z})]
    -(\alpha \tilde{\xi}_a)\text{sinh}(2z/\xi_a)\text{cosh}[\alpha(\tilde{L}-\tilde{z})]
    }
    {\alpha [(\alpha \tilde{\xi}_a)^2-4]}
    \nonumber
    \\
    &+ 
    \frac{-[2+(\beta \tilde{\xi}_a)^2] \text{sin}(\beta \tilde{L})
    +[2+(\beta \tilde{\xi}_a)^2\text{cosh}(z/\xi_a)^2]
    \text{sin}[\beta(\tilde{L}-\tilde{z})]
    -(\beta \tilde{\xi}_a)\text{sinh}(2z/\xi_a)\text{cos}[\beta(\tilde{L}-\tilde{z})]
    }
    {\beta [(\beta \tilde{\xi}_a)^2+4]},
    \\
    I_2(z,\Omega)=&
     \frac{
    [-2+(\alpha \tilde{\xi}_a)^2\text{cosh}[(L-z)/\xi_a]^2]
    \text{sinh}[\alpha(\tilde{L}-\tilde{z})]
    -(\alpha \tilde{\xi}_a)\text{sinh}[2(L-z)/\xi_a]\text{cosh}[\alpha(\tilde{L}-\tilde{z})]
    }
    {\alpha [(\alpha \tilde{\xi}_a)^2-4]}
    \nonumber
    \\
    &+ 
    \frac{
    -[2+(\beta \tilde{\xi}_a)^2\text{cosh}[(L-z)/\xi_a]^2]
    \text{sin}[\beta(\tilde{L}-\tilde{z})]
    -(\beta \tilde{\xi}_a)\text{sinh}[2(L-z)/\xi_a]\text{cos}[\beta(\tilde{L}-\tilde{z})]
    }
    {\beta [(\beta \tilde{\xi}_a)^2+4]}.
    \end{align}
    \end{widetext}
    
     \begin{figure*}[t]
    \includegraphics[width=1\textwidth]{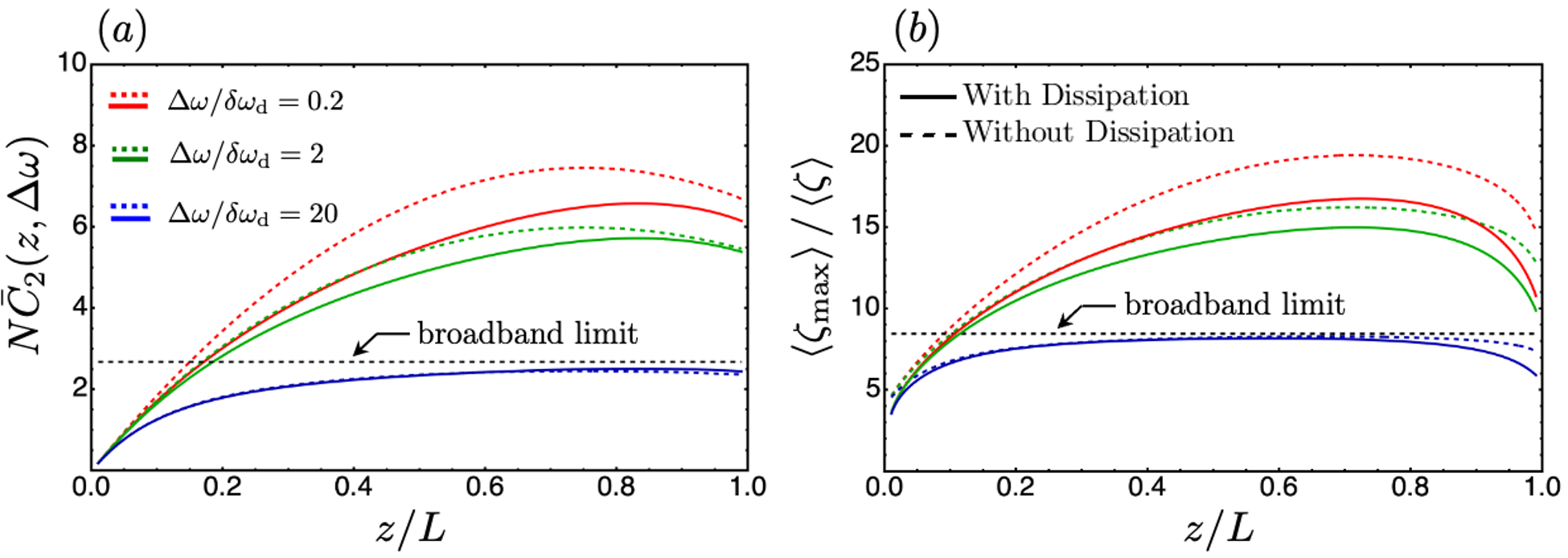}
    \caption{ {\bf Broadband long-range correlation and energy enhancement}. Theoretical predictions for the broadband long-range correlation $N\bar{C}_2(z,\Delta\omega) \simeq \text{Var}(\zeta)/\left< \zeta\right>^2$ in (a) and the maximum enhancement of energy deposition $\langle\zeta_{\text{max}} \rangle /\langle\zeta\rangle$ in (b), for different input bandwidths $\Delta \omega$ expressed in units of the diffusive correlation width $\delta\omega_d=1/\tau_d=\pi^2D/L^2$, with dissipation ($\xi_a/L=0.5$, solid lines) and without dissipation (dashed lines). Results in (a) are obtained by computing $\iint_{\Delta\omega}d\omega_1d\omega_2C_2^{(1)}(z,|\omega_1-\omega_2|)/\Delta\omega^2$ from Eq.~\eqref{EqC2Abs}, and the horizontal dashed line corresponds to the broadband approximation~\eqref{EqIntegratedC2Approx}. Results in (b) are obtained from the solution of Eq.~\eqref{EqResolvent3}, and the horizontal dashed line corresponds to the approximation~\eqref{EqEnhancement}, where $\gamma$ is replaced by its broadband expansion that follows from Eq.~\eqref{EqIntegratedC2Approx}, $\gamma=(8/\left< \tau\right>)L_{\Delta\omega}/L=(4\sqrt{2}/\pi\left< \tau\right>)\sqrt{\delta\omega_{d}/\Delta\omega}$. The average transmittance of the diffusive waveguide is $\left<\tau \right>=0.1$.
    }
    \label{FigPredictions}
    \end{figure*}
    
    The expression of $C_2(z,\Omega)$ can be considerably simplified in the broadband limit, when the conditions $L_\Omega \ll L, \xi_a$ and $z\gtrsim L_\Omega$ are satisfied, as is the case in our experiment. In this regime, we find that $C_2(z,\Omega)$ becomes independent of depth and dissipation,
    \be
    C_2(z,\Omega)\simeq \frac{2}{g}\frac{L_\Omega}{L}  \simeq \frac{4}{\pi} \frac{1}{N}\frac{L_\Omega}{\ell_t}.
    \label{EqC2Approx}
    \ee
    This result can be understood qualitatively with the following argument. In the monochromatic case, the crossing of diffusive paths and the exchange of field partners shown in Fig.~\ref{FigDiagrams} can occur at any depth throughout the sample, with a probability of $\sim 1/g$. In contrast, in the broadband case, the field correlation $\vert\langle \mathcal{I}(z,\Omega)\rangle\vert^2\sim e^{-z/L_\Omega}$ survives within a depth $\sim L_\Omega$, forcing the exchange of field partners to occur within this depth. The crossing probability is therefore reduced by a factor $\sim L_\Omega/L$. Dissipation only modifies this result for $\xi_a \lesssim  L_\Omega$. Using the simplified expression~\eqref{EqC2Approx}, we find that the long-range correlation integrated over the bandwidth $\Delta\omega$ can be approximated by
    \be
    \bar{C}_2(z,\Delta\omega)\simeq \frac{16}{3g}\frac{L_{\Delta\omega}}{L}  \simeq \frac{32}{3\pi} \frac{1}{N}\frac{L_{\Delta\omega}}{\ell_t},
    \label{EqIntegratedC2Approx}
    \ee
     for $L_{\Delta\omega} \ll L, \xi_a$ and $z\gtrsim L_{\Delta\omega}$. This broadband result is compared with the exact integration of Eq.~\eqref{EqC2Abs} in Fig.~\ref{FigPredictions}(a), with and without dissipation. In this limit, the parameter $\gamma$ controlling the value of the energy deposition enhancement in Eq.~\eqref{EqEnhancement} becomes $\gamma\simeq(16/\pi)L_{\Delta\omega}/\ell_t$. This estimate of the energy enhancement is compared with the exact result derived from Eq.~\eqref{EqResolvent3} in Fig.~\ref{FigPredictions}(b).

    \subsection{Focusing and broadband short-range correlations}
    
    \begin{figure*}
    	\includegraphics[width=\textwidth]{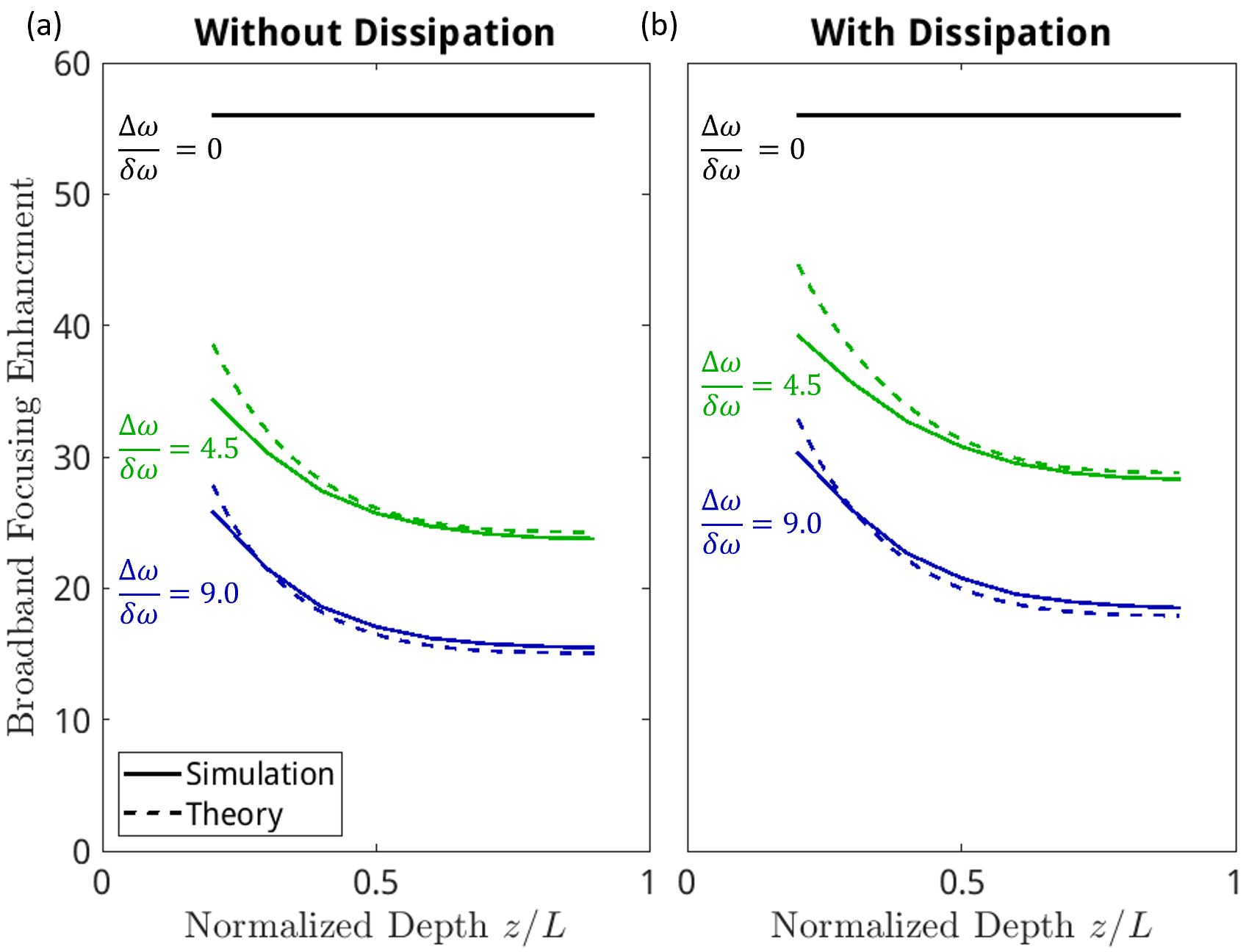}
    	\caption{\textbf{Broadband focusing to wavelength-scale target}. Broadband focusing enhancement as a function of normalized depth $z/L$ without dissipation (a) and with dissipation (b). Input bandwidth $\Delta\omega/\delta\omega$ = 0 (black), 4.5 (green), and 9.0 (purple), where $\delta\omega$ is the spectral correlation width calculated at $z/L = 3/5$. For monochromatic light, the focusing enhancement is equal to the number of input channels $N$ = 56 at any depth, with and without dissipation. For broadband input, focusing enhancement is smaller at larger depth. The decrease is more pronounced for larger bandwidth. Both numerical simulations (solid line) and analytical predictions from Eqs.~\eqref{EqEnhancementFocusing},~\eqref{EqMeffFocusing} and~\eqref{EqC1Exact} (dashed line) show that dissipation increases the broadband focusing enhancement at all depths. \label{Focus}}
    \end{figure*}
    
    Broadband focusing to a wavelength-scale speckle can still be modeled by a broadband deposition matrix $\mathcal{A}$ of the form given by Eq.~\eqref{EqMatA}, where the matrices $\mathcal{Z}(\omega)$ are of size $1\times N$. In this regime, these matrices behave as Gaussian random matrices~\cite{Goetschy2013}. This implies that the Stieltjes transform $g_{\mathcal{A}}(w)$ still obeys Eq.~\eqref{EqResolvent1} where $g_{\mathcal{Z}^\dagger\mathcal{Z}}(w)$ is the Stieltjes transform of the Wishart matrix $\mathcal{Z}^\dagger\mathcal{Z}$. The solution $g_{\mathcal{A}}(w)$ is identical to the Stieltjes transform of a new Wishart matrix of aspect ratio $N/M_{\text{eff}}$. In other words, the matrix $\mathcal{A}$ has the same eigenvalue spectrum as a matrix $X^\dagger X$, where $X$ is a $M_{\text{eff}}\times N$ Gaussian random matrix. For $N, M_{\text{eff}} \gg1$, the eigenvalue distribution satisfies the Marchenko-Pastur law, whose upper edge gives the focusing enhancement
    \be
    \frac{\langle\zeta_{\text{max}}\rangle}{\langle\zeta\rangle}=
    \left(1+\sqrt{\frac{N}{M_{\text{eff}}}} \right)^2.
    \label{EqEnhancementFocusing}
    \ee 
    
    In addition, the effective number of independent channels $M_{\text{eff}}$ is still given by Eq.~\eqref{EqMeffExact}, where the correlation function $C^{(T)}(z,\omega_1,\omega_2)$ is now dominated by its short-range component $C_1(z, \omega_1,\omega_2)$, because the matrices $\mathcal{Z}(\omega)$ possess a single output channel. As $C_1(z, \omega_0,\omega_0)=1$ at all depth, $M_{\text{eff}}$ is given by
    \be
    \frac{1}{M_{\text{eff}}}=\iint_{\Delta\omega}\frac{d\omega_1d\omega_2}{\Delta\omega^2}C_1(z,\omega_1,\omega_2),
    \label{EqMeffFocusing}
    \ee 
    where $C_1(z,,\omega,\omega +\Omega)\equiv C_1(z,\Omega)$ is given by
    \begin{widetext}
    \begin{align}
    C_1(z,\Omega)= \frac{\vert\left<E(z,\omega)E(z,\omega+\Omega)^* \right>\vert^2}{\left<E(z,\omega)E(z,\omega)^* \right>^2}=\frac{\vert\mathcal{I}(z,\Omega) \vert^2}{\mathcal{I}(z,0)^2}.
    \label{EqDefC1}
    \end{align}
    Using the expression of $\mathcal{I}(z,\Omega)$ discussed in the previous section, we find that
    \begin{align}
    C_1(z,\Omega)=\frac{\text{sinh}(L/\xi_a)^2}{\text{sinh}[(L-z)/\xi_a]^2}
    \frac
    {\text{cosh}[\alpha (\tilde{L}-\tilde{z})]-\text{cos}[\beta (\tilde{L}-\tilde{z})]}
    {\text{cosh}(\alpha \tilde{L})-\text{cos}(\beta \tilde{L})}.
    \label{EqC1Exact}
    \end{align}
    \end{widetext}
    In the regime $ L \gtrsim \xi_a \gg L_\Omega$ and $L-z\gtrsim L_\Omega$, it can be approximated as
    \be
    C_1(z,\Omega)\simeq e^{-z/L_\Omega}e^{2z/\xi_a}.
    \label{EqC1Aprox}
    \ee
    We note that $C_1(z,\Omega)$ increases with dissipation in the broadband limit, contrary to $C_2(z,\Omega)$ [see Eq.~\eqref{EqC2Approx}]. The reason is that the numerator of Eq.~\eqref{EqDefC1} behaves as $\sim e^{-z/L_\Omega}$, because it is made up of diffusive paths at frequency $\Omega$ that propagate over a distance $L_\Omega \ll \xi_a$, whereas the denominator is made up of stationary diffusive paths that propagate over a distance $\xi_a$ and scales as $e^{-2z/\xi_a}$. On the contrary, the numerator of $C_2(z,\Omega)$ is mainly made up of a crossing of paths followed by stationary diffusive paths whose contributions cancel out the dissipation dependence of the denominator [see Fig.~\ref{FigDiagrams}(a)].
    
    Using the approximate expression~\eqref{EqC1Aprox} in Eq.~\eqref{EqMeffFocusing}, we find the following broadband expansion of $M_{\text{eff}}$ deep inside the medium,
    \be
    \frac{1}{M_{\text{eff}}}\simeq \frac{4}{\bar{z}^2}\left[1-\frac{6}{\bar{z}^2} + 2 e^{-\bar{z}}\left(1+\frac{3}{\bar{z}}+\frac{3}{\bar{z}^2}\right)\right]e^{2z/\xi_a},
    \ee
    where $\bar{z}=z/L_{\Delta \omega}$. In the regime where $N \gg M_{\text{eff}}$, we conclude that the broadband focusing enhancement~\eqref{EqEnhancementFocusing} decays with depth $z$ as
    \be
    \frac{\langle\zeta_{\text{max}}\rangle}{\langle\zeta\rangle} \simeq \frac{N}{M_{\text{eff}}} \simeq 4N\left(\frac{L_{\Delta \omega}}{z}\right)^2
    \label{EqfocusEnhance}
    \ee
    in the absence of dissipation. 
    
    In figure \ref{Focus}, we compute the focusing enhancement analytically and compare with numerical simulation results, both with and without dissipation. The focusing enhancement decays rapidly with both increasing depth and input bandwidth, following Eq.~\eqref{EqfocusEnhance}. This is in stark contrast to the broadband energy deposition to extended targets, for which the depth dependence vanishes at large bandwidth [see Eq.~\eqref{EqEnhancement}]. In addition, dissipation effects increase the focusing enhancement as predicted by Eq. \eqref{EqC1Aprox}. This further contrasts broadband energy deposition to extended targets, for which dissipation dependence also vanishes at large bandwidth, as shown by Eq. \eqref{EqEnhancement}.

	\bibliography{main}
	
\end{document}